\documentclass[aps,prd,twocolumn,showpacs,floatfix,preprintnumbers,amsmath,amssymb,nofootinbib]{revtex4-1}
\usepackage{graphicx}

\usepackage[breaklinks, colorlinks, citecolor=blue, colorlinks=false, linkcolor=blue]{hyperref}

\newcommand{\mnras}{Mon. Not. R. Astron. Soc.}

\def\hatn{\mathbf{\hat n}}

\newcommand{\lsim}{\mathrel{\hbox{\rlap{\lower.55ex\hbox{$\sim$}} \kern-.3em \raise.4ex \hbox{$<$}}}}
\newcommand{\gsim}{\mathrel{\hbox{\rlap{\lower.55ex\hbox{$\sim$}} \kern-.3em \raise.4ex \hbox{$>$}}}}
\def\wigner#1#2#3#4#5#6{ \left( \begin{array}{ccc} #1 & #3 & #5
\\ #2 & #4 & #6 \\ \end{array} \right)}

\newcommand{\threej}[6]{\left(
    \begin{array}{ccc}
        \! #1\! & #2\!  & #3\!  \\
        \! #4\! & #5\!  & #6\!
      \end{array}
    \right)}
    
\newcommand{\sixj}[6]{\left\{
                      \begin{array}{ccc}
    #1 & #2  & #3  \\
    #4 & #5  & #6
                      \end{array}
               \right\}}
\def\be{\begin{equation}}
\def\ee{\end{equation}}
\def\bl{\mathbf{l}}
\def\norm{ {\cal N} }

\begin{document}
\title{Baryons do trace dark matter 380,000 years after the big bang:\\Search for compensated isocurvature perturbations with WMAP 9-year data}
\author{Daniel Grin$^1$, Duncan Hanson$^{2,3}$, Gilbert P. Holder$^2$, Olivier Dor\'e$^{3,4}$, and Marc Kamionkowski$^5$}
\affiliation{$^1$School of Natural Sciences, Institute for Advanced
     Study, Princeton, NJ 08540}
\affiliation{$^2$Department of Physics, McGill University, Montreal QC H3A 2T8, Canada}
\affiliation{$^3$Jet Propulsion Laboratory, California Institute
     of Technology, Pasadena, CA 91109}
\affiliation{$^4$California Institute of Technology, Pasadena, CA 91125} 
\affiliation{$^5$Department of Physics and Astronomy, Johns Hopkins University, Baltimore, MD 21218, USA}

\date{\today}
\begin{abstract}
Primordial isocurvature fluctuations between photons and either neutrinos or non-relativistic species such as baryons or dark matter are known to be sub-dominant to adiabatic
fluctuations. Perturbations in the relative densities of baryons and dark matter (known as compensated isocurvature perturbations, or CIPs), however, are surprisingly poorly constrained. CIPs leave no imprint in the cosmic microwave background (CMB) on observable scales, at least at linear order in their amplitude and zeroth order in the amplitude of adiabatic perturbations. It is thus not yet empirically known if baryons trace dark matter at the surface of last scattering. If CIPs exist, they would spatially modulate the Silk damping scale and acoustic horizon, causing distinct fluctuations in the CMB temperature/polarization power spectra across the sky: this effect is first order in both the CIP and adiabatic mode amplitudes. Here, temperature data from the Wilkinson Microwave Anisotropy Probe (WMAP) are used to conduct the first CMB-based observational search for CIPs, using off-diagonal correlations and the CMB trispectrum. Reconstruction noise from weak lensing and point sources is shown to be negligible for this data set. No evidence for CIPs is observed, and a $95\%$-confidence upper limit of $1.1\times 10^{-2}$ is imposed to the amplitude of a scale-invariant CIP power spectrum. This limit agrees with CIP sensitivity forecasts for WMAP, and is competitive with smaller scale constraints from measurements of the baryon fraction in galaxy clusters. It is shown that the root-mean-squared CIP amplitude on $5-100^{\circ}$ scales is smaller than $\sim0.07-0.17$ (depending on the scale) at the $95\%$-confidence level. Temperature data from the \textit{Planck} satellite will provide an even more sensitive probe for the existence of CIPs, as will the upcoming ACTPol and SPTPol experiments on smaller angular scales. 
\end{abstract}
\maketitle


\section{Introduction}\label{iref}
Measurements of primordial density perturbations are consistent with adiabatic initial conditions, for which the ratios of neutrino, photon, baryon, and dark-matter number densities are initially spatially constant. The simplest inflationary models predict adiabatic fluctuations \cite{Guth:1982ec,Linde:1982uu,bardeen_iso,Hawking:1982cz,mukhanov_iso,starobinsky_iso}. Isocurvature perturbations, on the other hand, are fluctuations in the relative number densities of different species. They are produced in topological-defect models for structure formation
\cite{branden_topo}, multi-field inflationary models, curvaton models \cite{linde_iso,linde_mukhanov,mukhanov_b,langlois_riazuelo,langlois_double_iso}, in which two different fields drive inflation and generate curvature perturbations, and in simple inflationary models if the dark matter is composed of axions \cite{axenides,seckel_turner,linde_iso}.

CMB temperature anisotropies limit the contribution of baryon
isocurvature perturbations (fluctuations in the
baryon-to-photon ratio) \cite{peebles_iso_a,peebles_iso_b} and CDM
isocurvature perturbations (fluctuations in the
dark-matter--to--photon ratio) \cite{burns_axion,seckel_turner,hu_iso_a,hu_iso_b} to the total perturbation amplitude \cite{boomerang_iso_limits,planck_iso,wmap7_komatsu_a,wmap7_komatsu_b,pl_inflation,tommaso,zunckel_iso,bucher_dunkley_a,kawasaki_iso,beltran_viel_iso,seljak_iso,riazuelo_iso_constraint,takahashi_iso_nongauss}. The recent \textit{Planck} CMB results limit the CDM isocurvature fraction to be $\lesssim 3.9\%$ of the total perturbation amplitude.

It is therefore surprising that perturbations in the baryon
density can be almost arbitrarily large, as long they are
compensated by dark-matter perturbations such that the total-nonrelativistic-matter density
remains unchanged \cite{gordon_pritchard,holder}. These compensated isocurvature perturbations (CIPs) obey \begin{eqnarray} \rho_{\rm c}\delta_{\rm c}^{\rm CI}+\rho_{\rm b}\delta^{\rm CI}_{\rm b}=0,~~~\delta_{\gamma}^{\rm CI}=0,\end{eqnarray} where $\delta_{\rm c}$, $\delta_{\rm b}$, and $\delta_{\rm \gamma}$ are fractional energy density perturbations in the cold dark matter, baryons, and photons, respectively, while $\rho_{\rm c}$ and $\rho_{\rm b}$ are the homogeneous dark matter and baryon densities. CIPs induce no curvature perturbation at early times, and they therefore leave the photon density---and thus large-angle CMB
fluctuations---homogeneous at linear order.  

Curvaton models for inflation may generate CIPs
\cite{lyth_ungarelli_wands,gupta_malik_wands,gordon_lewis_curvaton,enqvist_subdom_curvaton},
with amplitudes approaching the regime detectable by the proposed EPIC mission
\cite{gordon_pritchard}, and other inflationary models \cite{spolyar} could generate even larger CIP amplitudes.
Recent theoretical ideas
\cite{Kaplan:2009ag,buckley_hylo,cladogenesis,wimp_baryo,heckman_dm,barymorphosis}
connecting the baryon asymmetry and dark-matter density could also have implications for CIPs. In any case, our principal motivation in studying CIPs is curiosity: can we
determine empirically, rather than simply assume, that the primordial
baryon fraction is homogeneous and traces the dark matter?

CIPs induce baryon motion through baryon-pressure gradients, but these motions occur only at the baryon sound speed. The resulting anisotropies would be imprinted on the baryonic sound horizon,
at $l\sim 10^6$ \cite{gordon_pritchard,gordon_lewis_curvaton,challinor_compensate}. Existing measurements at low redshift 
constrain the CIP perturbation amplitude to be $\lesssim10\%$ \cite{holder,gordon_pritchard}, while more sensitive proposed measurements of 21-cm absorption during the cosmic dark ages are a way off in the future \cite{barkana,challinor_compensate,gordon_pritchard,sekigu}.

More recently, it has been shown that CIPs would modulate the CMB anisotropies produced by adiabatic perturbations, both by inducing anisotropies in the optical depth to reionization \cite{holder}, and more dramatically, by changing the Silk damping length of the CMB in regions of sky containing a CIP \cite{grin_prl,grin_prd}. This modulation would induce a specific pattern of higher order-correlations in the temperature and polarization anisotropies, analogous to those induced by variations of other cosmological parameters \cite{sigurdson_alpha}, and by weak gravitational lensing \cite{lewis_lens_review}.
This signature can be exploited to construct estimators for the CIP perturbation \cite{grin_prd}.
CIPs generated in curvaton models for the primordial density fluctuation \cite{lyth_ungarelli_wands,gupta_malik_wands,gordon_lewis_curvaton,enqvist_subdom_curvaton}, lower-energy inflationary models \cite{spolyar}, and perhaps other scenarios, are within the range of detectability for a cosmic-variance-limited CMB polarization experiment. Our motivation for studying CIPs, however is here one of curiosity: If pre-existing limits are so permissive, it behooves us to actually check if baryons and dark matter trace one another in the early universe! 

Here, we use WMAP 9-year temperature maps to search for CIPs.
Our CIP estimator is based on the full non-Gaussian trispectrum of the observed CMB multipole moments, and can be used to perform either a model-independent reconstruction of the CIP power spectrum $C_{L}^{\Delta \Delta}$, or to measure the amplitude of a scale-invariant spectrum of CIPs. 

We impose a $95\%$-confidence upper limit of $1.1\times 10^{-2}$ to the amplitude of a scale-invariant spectrum of CIPs, as well as model-independent constraints of $\sim 10\%$ to the root-mean-squared (RMS) amplitude of the CIP power spectrum at angular scales in the range $1\leq L<20$, where $L$ is the multipole index of the CIP. We show that secondary contractions of the trispectrum contribute negligibly to the estimator, at least for WMAP experimental parameters. We show that known sources of non-Gaussianity, such as gravitational lensing and unresolved point sources, do not provide a significant bias for our estimates of the CIP power spectrum. The same methodology which we have used here could be applied to the \textit{Planck} data, and has the potential to significantly improve on the constraints above.

We begin in Sec. \ref{cip_rederive} with a derivation of the temperature anisotropies induced by CIPs. In Sec. \ref{esec}, we present our CIP estimator, based on that of Ref. \cite{grin_prd}, but generalized to be run on a partial sky-map with realistic noise properties.  We account for bias and estimator normalization using analytic estimates and Monte Carlo simulations. 
In Sec. \ref{bam} we present our results and compare with the forecasted WMAP sensitivity, and we conclude in Sec. \ref{conclusions}. 
Throughout this work we use a fixed, flat $\Lambda$CDM cosmology consistent with the WMAP-9 \cite{Hinshaw:2012fq} power spectrum, given by
$\Omega_b = 0.045$, $\Omega_c = 0.222$, $h = 0.733$, adiabatic spectral index $n_s = 0.963$, reionization optical depth $\tau = 0.088$, and adiabatic scalar power spectrum normalization $A_s = 2.4 \times 10^{-9}$.

\section{CMB temperature correlations in the presence of a CIP}
In Refs. \cite{grin_prd} and \cite{grin_prl}, it was shown that a CIP would induce off-diagonal correlations between CMB anisotropy multipole moments. We rederive these results using a clearer method here, applying the line-of-sight (LOS) formalism of Ref. \cite{matias_uros_los}. This approach is more readily generalized to CIPs with wavelength smaller than the thickness of the surface of last-scattering, and is useful in computing the CIP bispectrum. Before launching into the formalism, we review the physical origin of CIP-induced CMB correlations.

As noted in Refs. \cite{gordon_pritchard,gordon_lewis_curvaton,challinor_compensate}, CIPs have no initial potential perturbations (like other isocurvature modes), but also have no initial radiation pressure gradients. In linear theory, flows then begin at the baryon sound speed, but are only effective in transferring fluctuations to the photons on very small scales ($l\sim 10^{6}$). Even when baryons do begin to evacuate initial density fluctuations (which would yield a net potential perturbation as the CIP evolved), these flows will be diffusion damped, slowing down the growth of these small-scale potential perturbations. This intuition is confirmed by running the Boltzmann code \textsc{camb} \cite{camb} with a CIP initial condition, and noting that the induced CMB temperature anisotropy is negligible. 

The fluctuating baryon fraction in the presence of a CIP would lead to an inhomogeneous redshift of reionization (when the first sources turn on), leading to a fluctuating optical depth and a distinct patchy reionization signal in the CMB \cite{holder}. More dramatically, however, if CIPs are in fact primordial, they will lead to a spatial modulation of coefficients in the early-time (tight-coupling era) equations of motion for fluid perturbations, altering observed CMB anisotropies in a detectable way \cite{grin_prd,grin_prl}.

\label{cip_rederive}
\subsection{Physical origin of effect}
\begin{figure*}[htbp]
\includegraphics[width=6.50in]{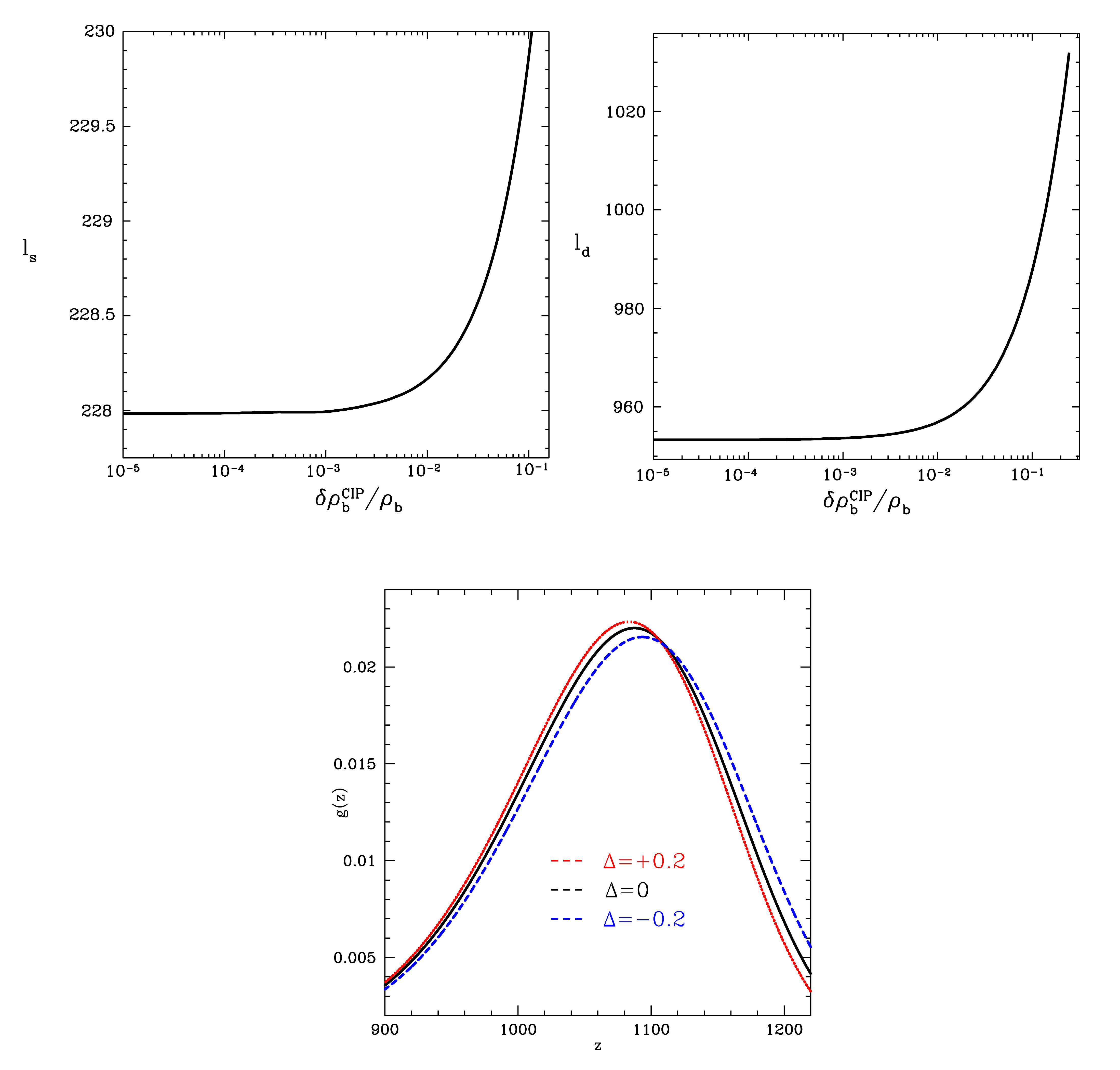}
\caption{Physical and Thomson scattering visibility function $g(z)$ in the presence of a global CIP $\Delta$. Top left panel shows angular sound horizon $l_{\rm s}$ as a function of of a spatially uniform CIP $\Delta$. Top right panel shows diffusion damping scale $l_{\rm d}$ as a function of $\Delta$. Bottom panel shows $g(z)$ evaluated for $3$ different values of $\Delta$. Later we will use these physical effects to probe the CMB for spatially varying $\Delta(\hat{n})$.}
\label{intuit}
\end{figure*}

Consider a compensated isocurvature perturbation (CIP) along the line of sight $\Delta(\hat{n})$, which gives local changes in the baryon and CDM energy densities given by
\begin{eqnarray}
\Omega_b &\rightarrow \Omega_b[1+\Delta(\hat{n})], \nonumber \\
\Omega_c &\rightarrow \Omega_c - \Omega_b \Delta(\hat{n}).
\end{eqnarray}
A positive (negative) value of $\Delta$ corresponds to a more (less) baryon-loaded plasma, which decreases (increases) the sound speed and thus decreases (increases) the physical acoustic horizon. The multipole index $l_{s}$ of the first CMB acoustic peak thus increases (decreases) as $\Delta$ increases (decreases), as shown in Fig. \ref{intuit} (from Ref. \cite{grin_prd}), generated using expressions in Ref. \cite{dodelcosmo}. The other acoustic peak locations behave similarly.

CMB temperature anisotropies are suppressed on angular scales $l>l_{\rm d}\sim 1000$ due to diffusion damping. Using the expressions in Ref. \cite{matias_harari} and the \textsc{camb} \cite{camb} code, we evaluate $l_{d}(\Delta)$ and show the results in the top right panel of Fig. \ref{intuit}. We see that, as photons diffuse over smaller distances, as a result of higher local baryon density in the presence of a CIP with positive $\Delta$, the transition to exponential damping of CMB anisotropies occurs at higher $l$.

In the bottom panel of Fig. \ref{intuit} (from Ref. \cite{grin_prd}), we show the visibility functions $g(z)=e^{-\tau}d\tau/dz$ for $3$ different values of $\Delta$; $\tau$ is the optical depth due to Thomson scattering and $z$ is the redshift. The peak of the visibility function is the redshift $z_{\rm SLS}$, at which most CMB photons last scatter. In the presence of a positive (negative) $\Delta$ CIP, decoupling occurs later (earlier) due to higher (lower) baryon density. 

The effects described above all result in modifications to the CMB power spectrum.
One can therefore imagine constructing an estimator for spatial variations in $\Delta$ by forming localized power spectrum estimates and fitting for $\Delta$ relative to a fiducial model based on the full-sky power spectrum. 
We will do this in Sec.~\ref{esec}, however first we must quantify the effect of CIPs on the local power spectrum, which we do using the LOS formalism in the following sections.
\subsection{Standard line-of-sight solution for CMB temperature anisotropy}
The calculation of CMB anisotropies is greatly simplified using the line-of-sight (LOS) approach, introduced in Ref. \cite{matias_uros_los}. At spatial location $\vec{y}$, conformal time $\eta$, and for photon momentum direction vector $\hat{p}$ (where $|\hat{p}|=1$), the photon temperature perturbation $T(\vec{y},\hat{p},\eta)$ obeys the equation (derived from the Boltzmann equation)
\begin{equation}
\dot{T}(\vec{y},\hat{p},\eta)+\hat{p}\cdot \nabla T(\vec{y},\hat{p},\eta)=\mathcal{D}(\hat{p},\eta)[\vec{u}(\vec{y},\eta)],
\end{equation}where $\vec{u}(\vec{y},\eta)$ is a vector whose entries are the fluid density/velocity perturbations (as well as higher-order moments of the distribution function, for neutrinos) and metric fluctuations characterizing the system, and $\mathcal{D}$ is a linear differential operator which maps $\vec{u}(\vec{y},\eta)$ to a source term for temperature perturbations. Defining the LHS of this equation as the operator $\mathcal{B}$ (for Boltzmann), and taking a Fourier transform, we obtain
\begin{equation}
\mathcal{B}_{\vec{k}}[T_{\vec{k}}(\hat{p},\eta)]=\mathcal{D}_{\vec{k}}(\hat{p},\eta)[\vec{u}_{\vec{k}}(\eta)],\label{evo_eq}
\end{equation}where $\mathcal{D}_{\vec{k}}$ is a matrix operator in Fourier space (as opposed to a differential operator). For the usual adiabatic mode, $\vec{u}_{\vec{k}}(\eta)=\vec{f}_{\vec{k}}(\eta)\Phi_{\vec{k}}$, where $\vec{f}_{\vec{k}}(\eta)$ is a time evolution operator mapping the initial potential perturbation $\Phi_{\vec{k}}$ to the solution for the fluid variables at subsequent times. 

The components of the matrix operator $\mathcal{D}_{\vec{k}}$ (which maps the fluid and metric variables to the observed temperature perturbation) and vector-valued function $\vec{f}_{\vec{k}}$ are laid out in detail in Refs. \cite{matias_uros_los,ma_bertschinger} and others, so we use the operator notation to keep things simple and general. Eq.~(\ref{evo_eq}) may be formally integrated to obtain 
\begin{equation}
T_{\vec{k}}(\hat{p},\eta_{0})=\int_{0}^{\eta_{0}}d\eta e^{ik\mu\left(\eta_{0}-\eta\right)}\tilde{S}[\hat{p},\vec{k},\eta]\Phi_{\vec{k}}\label{befit}\end{equation} where the source function is $\tilde{S}[\hat{p},\vec{k},\eta]=\mathcal{D}_{\vec{k}}(\hat{p},\eta)\vec{f}_{\vec{k}}(\eta)$, $\eta_{0}$ is the conformal time today and $\mu=\hat{p}\cdot\vec{k}/|\vec{k}|$. It turns out that the source function depends on $\vec{k}$ and $\hat{p}$ only through $\mu$ (as a polynomial in $\mu$) and $k=|\vec{k}|$, and so Eq.~(\ref{befit}) may be integrated by parts to obtain
\begin{equation}
T_{\vec{k}}(\hat{p},\eta_{0})=\int_{0}^{\eta_{0}}d\eta e^{ik\mu\left(\eta_{0}-\eta\right)}S[k,\eta]\Phi_{\vec{k}},
\end{equation}in terms of a different source function $S[k,\eta]$, whose terms are specified in Refs. \cite{dodelcosmo,matias_uros_los}. Going back to real space, following suit with an inverse spherical harmonic transform to derive multipole coefficients, and expanding the exponential using a Fourier-Bessel series, we obtain
\begin{eqnarray}
T_{lm}=\frac{4\pi i^{l}}{\left(2\pi\right)^{3}}\int d^{3}k T_{l}(k)\Phi_{\vec{k}}Y_{lm}^{*}(\hat{k}),\label{loss}\\
T_{l}(k)\equiv \int_{0}^{\eta_{0}}S[k,\eta]j_{l}[k(\eta_{0}-\eta)]
\end{eqnarray}
where $j_{l}(x)$ is a spherical Bessel function of index $l$. 

The familiar CMB angular power spectrum may then be obtained:
\begin{eqnarray}
\left \langle T_{lm}T^{*}_{l'm'}\right \rangle&=&\delta_{l l'}\delta_{mm'}C_{l},\\
C_l&=&\frac{2}{\pi}\int k^{2}dk T_{l}^{2}(k)P_{\Phi}(k),
\end{eqnarray}where the $3$-dimensional potential fluctuation power spectrum $P_{\Phi}(k)$ is defined by
\begin{equation}
\left \langle \Phi_{\vec{k}} \Phi^{*}_{\vec{k'}}\right \rangle=\left(2\pi\right)^{3}\delta^{3}(\vec{k}-\vec{k'})P_{\Phi}(k).\end{equation}
We now generalize the LOS solution to compute the off-diagonal temperature correlations induced by a CIP.
\\ \\
\subsection{Line-of-sight solution in the presence of a CIP}
The operators $\mathcal{D}_{\vec{k}}(\eta)$ and $\vec{f}_{\vec{k}}(\eta)$ depend on the cosmological parameters and thus also on the amplitude $\Delta(\hat{n})$ of a CIP. In the presence of a CIP, the real-space evolution equation will read
\begin{equation}
\mathcal{B}[T(\vec{y},\hat{p},\eta)]=\frac{d\mathcal{D}(\hat{p},\eta)[\vec{u}(\vec{y},\eta)]}{d\Delta}\Delta(\vec{y}).\label{modulated}
\end{equation}

As discussed above, on the scales of interest, the CIP amplitude is frozen in time. In principle, it still has dependence on the conformal time along a photon trajectory, if the mode wavelength is shorter than the integration interval. For modes of large angular scale $l\ll l_{\rm silk}$, however, this radial dependence may be neglected, and $\Delta(\vec{y})=\Delta(\hat{y})$. Fourier transforming Eq.~(\ref{modulated}), performing a spherical harmonic expansion of $\Delta(\hat{y})=\sum_{LM}\Delta_{LM}Y_{LM}(\hat{n})$, we see that the evolution equation for $T_{\vec{k}}(\hat{p},\eta)$ is of the same form as Eq.~(\ref{evo_eq}), but with a source term that is a linear superposition of source terms like this in Eq.~(\ref{evo_eq}). Since this is a linear system, solutions may be superimposed. Fourier-Bessel expanding and performing an inverse spherical harmonic transform of the solution, we obtain the perturbation to the LOS solution induced by a  CIP:\begin{widetext}
\begin{align}
\delta T_{lm}=&\sum_{LM l_1 m_1}\frac{\Delta_{LM}4\pi i^{l_{1}}\xi_{lm l_1 m_1}^{LM}K^{L}_{l l_{1}}}{\left(2\pi\right)^{3}}\int d^{3} qY_{l_{1}m_{1}}^{*} (\hat{q})\Phi_{\vec{q}}\frac{dT_{l_{2}}(q)}{d\Delta},\label{induced}\\
 \xi^{LM}_{lml_{1}m_{1}}  \equiv &
     {\left(K^{L}_{ll_{1}}\right)}^{-1} \int d\hat{n}Y^{*}_{lm}
     (\hat{n}) Y_{LM}(\hat{n})
     Y_{l_{1}m_{1}} (\hat{n}) 
      = \left(-1\right)^{m} \sqrt{\frac{\left(2L+1\right)
      \left(2l+1\right) \left(2l_{1}+1\right)}{4\pi}} 
  \times \wigner{l}{-m}{L}{M}{l_{1}}{m_{1}},\label{xideff}\\\nonumber
   ~~~    K^{L}_{ll_{1}}\equiv&\wigner{l}{0}{L}{0}{l_{1}}{0},
\end{align}
\end{widetext}expressed in terms of the familiar Wigner-3J symbols \cite{angmom}.

\subsection{CIP Statistics}
As discussed in the previous section, a compensated isocurvature perturbation induces a small fluctuation in the CMB temperature which is proportional to the primordial potential $\Phi_{\vec{k}}$.
If we consider a fixed realization of isocurvature perturbations $\Delta_{LM}$, the effect of CIPs is to introduce ``statistical anisotropy'' into the CMB, which is manifest as off-diagonal elements in the covariance matrix of the CMB fluctuations.
These are given by
\begin{multline}
     \langle T^{*}_{l'm'}T_{lm} \rangle =
     \delta_{l l'} \delta_{m m'}
     C_{l}^{\rm
     TT} \\ + \sum_{LM}
     \threej{l}{l'}{L}{m}{m'}{M} W^{\Delta}_{l l' L} \Delta_{LM}.
\end{multline}
Here $C_{l}^{\rm TT}$ is the usual CIP-free temperature power spectrum.
The quantity
\begin{multline}
    W^{\Delta}_{l l' L} \equiv
    \sqrt{ \frac{(2l+1)(2l'+1)(2L+1)}{4\pi}} \\ \times
     \left(C_{l'}^{\rm T,dT}+C_{l}^{\rm
     T,dT}\right) \threej{l}{l'}{L}{0}{0}{0},
\label{toffdia_middle}
\end{multline}
is a weight function associated with the CIPs and
\begin{equation}
     C_{l}^{\rm T, dT} \equiv \frac{2}{\pi} \int k^{2}
     dk\, P_{\Phi}(k) T_{l}(k) \frac{dT_{l}(k)}{d\Delta}. \label{toffdia_b}
\end{equation}
The derivative power spectra of Eq.~(\ref{toffdia_b}) are evaluated using the \textsc{camb} code, using numerical methods described in Ref. \cite{camb}, including a spatial modulation in the optical depth to reionization, $\tau$, as well as the much larger effect from physics near recombination.

We see above that a fixed CIP realization breaks the diagonality of the covariance matrix of the $T_{lm}$ in a very specific way, 
yielding a unique statistical signature which can be used to reconstruct the $\Delta_{LM}$ realization.
A similar derivation can reproduce the off-diagonal polarization correlations obtained in Ref. \cite{grin_prd}.

Of course, in reality we do not have a fixed CIP realization to observe. 
If the Universe as a whole has no preferred orientation then the CIP perturbations $\Delta_{LM}$ are themselves random variables, with some statistically isotropic distribution.
If we assume that $\Delta_{LM}$ are independent of the primordial fluctuations and Gaussian, then they are completely characterized by their power spectrum $C_L^{\Delta \Delta}$.
In this more realistic picture, the first distinctive statistical signature of CIPs appears in the CMB ``trispectrum'', or connected 4-point function.
The connected part of the 4-point function is zero for purely Gaussian fluctuations, but in the presence of CIPs it becomes non-zero.
Following Ref. \cite{Hu:2001fa}, the connected 4-point function must take the form
\begin{multline}
\langle {T}_{l_1 m_1} {T}_{l_2 m_2} {T}_{l_3 m_3} {T}_{l_4 m_4} \rangle_C 
= \\ \sum_{LM} (-1)^{M} 
T^{l_1 l_2}_{l_3 l_4} (L)  {\cal G}^{l_1 m_1 l_2 m_2 L}_{l_3 m_3 l_4 m_4 M},
\end{multline}
where $T^{l_1 l_2}_{l_3 l_4} (L)$ is known as the trispectrum, and 
we have used the Glebsch-Morgan coefficient
\be
{\cal G}^{l_1 m_1 l_2 m_2 L}_{l_3 m_3 l_4 m_4 M} = 
\threej{l_1}{l_2}{L}{m_1}{m_2}{-M}
\threej{l_3}{l_4}{L}{m_3}{m_4}{M}.
\ee
Symmetry of the four multipoles further requires that the trispectrum may be encoded as
\begin{multline}
T^{l_1 l_2}_{l_3 l_4} (L)  = 
P^{l_1 l_2}_{l_3 l_4} (L) + 
(2L+1) \sum_{L'} \Bigg[ \\
(-1)^{l_2 + l_3} \sixj{l_1}{l_2}{L}{l_4}{l_3}{L'} P^{l_1 l_3}_{l_2 l_4} (L') \\
+ (-1)^{L + L'} \sixj{l_1}{l_2}{L}{l_4}{l_3}{L'} P^{l_1 l_4}_{l_3 l_2} (L') \Bigg].\label{bigtri}
\end{multline}
The first term $P^{l_1 l_2}_{l_3 l_4} (L)$ is referred to as the primary contraction of the trispectrum, 
while the final two terms are known as secondary contractions.
The primary contraction introduced by CIPs is given simply by
\be
P^{l_1 l_2}_{l_3 l_4} (L) =
C_L^{\Delta \Delta} W^{\Delta}_{l_1 l_2 L} W^{\Delta}_{l_3 l_4 L}.
\label{eqn:cip_primary_contraction}
\ee

Observationally, we will see that the picture of statistical anisotropy introduced initially above is still a useful ansatz; as the Universe shows us only a single realization of both $\Phi_{\vec{k}}$ and $\Delta_{LM}$, a fixed $\Delta_{LM}$ model is experimentally indistinguishable from one in which the $\Delta_{LM}$ are Gaussian random variables \cite{Ferreira:1997wd}.

\section{CIP estimator}
\label{esec}
We can construct an estimator for the CIP power spectrum utilizing the connected trispectrum which it induces.
If we are interested in a single mode of this power spectrum $C_L^{\Delta \Delta}$,
then following Ref. \cite{Regan:2010cn}, for example, an Edgeworth expansion of the CMB likelihood leads to the following optimal (minimum-variance) trispectrum estimator
\begin{multline}
\hat{C}_L^{\Delta \Delta}
= 
\frac{\norm_L}{2L+1}
\sum_{M} \sum_{l_i m_i} (-1)^M
C_L^{\Delta \Delta, ({\rm fid})} {\cal G}^{l_1 m_1 l_2 m_2 L}_{l_3 m_3 l_4 m_4 M}
\\ \times
W^{\Delta}_{l_1 l_2 L} W^{\Delta}_{l_3 l_4 L}
\Big( 
\bar{T}_{l_1 m_1} \bar{T}_{l_2 m_2}
\bar{T}_{l_3 m_3} \bar{T}_{l_4 m_4}
\\ - \left[ 2 \bar{T}_{l_1 m_1} \bar{T}_{l_2 m_2} -  {\bar C}^{TT}_{l_1 m_1, l_2 m_2} \right] {\bar C}^{TT}_{l_3 m_3, l_4 m_4}
\\ - \left[ 2 \bar{T}_{l_1 m_1} \bar{T}_{l_3 m_3} -  {\bar C}^{TT}_{l_1 m_1, l_3 m_3} \right] {\bar C}^{TT}_{l_2 m_2, l_4 m_4}
\\ - \left[ 2 \bar{T}_{l_1 m_1} \bar{T}_{l_4 m_4} -  {\bar C}^{TT}_{l_1 m_1, l_4 m_4} \right] {\bar C}^{TT}_{l_2 m_2, l_3 m_3}
\Big),
\label{eqn:triest}
\end{multline}
where $\norm_L$ is a normalization, $\bar{T}_{lm}$ are a set of so called ``inverse-variance filtered'' multipoles determined from the data map (we will discuss how these are obtained in more detail shortly in Sec.~\ref{sec:estimator:filtering}), and $\bar{C}_{lm, l'm'}$ is their covariance matrix.

For any modern CMB experiment, which measures thousands of modes, the covariance matrices $\bar{C}_{lm, l'm'}$ have trillions of elements and are impossible to work with directly.
It is, however, computationally tractable to evaluate the expression above (which distills these covariance matrices down to a single number) using Monte Carlo simulations.
We will rewrite Eq.~\eqref{eqn:triest} in a form which makes the details of this evaluation clearer, and also makes connection to the discussion of statistical anisotropy in the previous section.

We begin by introducing the ``quadratic estimator'' $\bar{\Delta}_{LM}$, 
which is a function of two inverse-variance filtered temperature multipoles as
\be
\bar{\Delta}_{LM}[ \bar{T}_{lm}^{(1)}, \bar{T}_{lm}^{(2)} ] = 
\sum_{lm, l'm'} \threej{l}{l'}{L}{m}{m'}{M} 
W^{\Delta}_{l l' L} \bar{T}_{lm}^{(1)} \bar{T}_{l'm'}^{(2)}.
\ee
If we were considering the CIP realization to be fixed, this quadratic estimator is precisely the quantity which is required for optimal estimation of $\Delta_{LM}$, following the formalism of Quadratic Maximum Likelihood (QML) estimators \cite{Hirata:2002jy,kamion_derotate,pullen_kamion,dvorkin_smith_a,dvorkin_smith_b,gluscevic_kamion_fullsky,gluscald,grin_prl,grin_prd,Hanson:2009gu}. This estimator is related to the actual CIP multipole moment $\Delta_{LM}$ by an overall normalization.
We will often find it useful to work with such un-normalized quantities, which we will denote with overbars. The reason is that for these maximum-likelihood estimators, the normalization is formed from the inverse of the estimator Fisher matrix. For this reason, we have denoted the un-normalized estimator $\hat{\Delta}_{LM}$ with an overbear, in analogy to the inverse variance filtered temperature multipoles. This quantity is also the variance of the estimator, and so these un-normalized estimates are effectively inverse-noise weighted. This makes the un-normalized estimators useful for estimating other parameters, such as the overall amplitude of a scale-invariant spectrum of CIPs, as we shall see below. We note that $\Delta_{LM}$ may be evaluated rapidly (with computational cost ${\cal O}(l_{\rm max}^3)$ using fast spherical harmonic transforms) in position space as the product of two filtered maps
\begin{multline}
\bar{\Delta}_{LM}[ \bar{T}_{lm}^{(1)}, \bar{T}_{lm}^{(2)} ] = \\
\int d\hat{n} Y_{LM}^{*}(\hat{n})
\bar{T}^{(1)}(\hat{n})  S^{(2)}(\hat{n}) + [ (1) \leftrightarrow (2) ],
\end{multline}
where the filtered maps themselves are given by
\begin{align}
\bar{T}^{(a)}(\hat{n})=&\sum_{lm}Y_{lm}(\hat{n}) \bar{T}^{(a)}_{lm},\\
S^{(a)}(\hat{n})=&\sum_{lm}Y_{lm}(\hat{n}) C_{l}^{\rm T,dT} \bar{T}^{(a)}_{lm},\label{fast}
\end{align}where $a=1$ or $a=2$ as appropriate.

Using this notation, we rewrite the $C_L^{\Delta \Delta}$ estimator as
$\hat{C}_L^{\Delta \Delta} = \bar{C}_L^{\Delta \Delta} / \norm_L$,
where
\be
\bar{C}_L^{\Delta \Delta} = C_L^{\bar{\Delta} \bar{\Delta}} - D_L^{\bar{\Delta} \bar{\Delta} }.
\label{eqn:clpphat}
\ee
The naive \textit{un-normalized} power spectrum estimate $C_L^{\bar{\Delta} \bar{\Delta}}$ is given by
\begin{multline}
C_L^{\bar{\Delta} \bar{\Delta}} = 
\sum_{l_i m_i} \sum_{M} \frac{1}{(2L+1)}
\Bigg< \\
\left( \bar{\Delta}_{LM}[ \bar{T}_{lm}, \bar{T}_{lm} ] - \bar{\Delta}_{LM}[ \bar{T}_{lm}^{(g)}, \bar{T}_{lm}^{(g)} ]  \right)^* \\
\times \left( \bar{\Delta}_{LM}[ \bar{T}_{lm}, \bar{T}_{lm} ] - \bar{\Delta}_{LM}[ \bar{T}_{lm}^{(f)}, \bar{T}_{lm}^{(f)} ]  \right) \Bigg>_{g,f},
\label{eqn:cl}
 \end{multline}
 and the ``disconnected noise bias'' estimate is given by
\begin{multline}
D_L^{\bar{\Delta} \bar{\Delta}} = 
\sum_{l_i m_i} \sum_{M} \frac{1}{(2L+1)}
\Bigg< \\
- 4 \bar{\Delta}_{LM}[ \bar{T}_{lm}, \bar{T}_{lm}^{(g)} ] \bar{\Delta}_{LM}[ \bar{T}_{lm}, \bar{T}_{lm}^{(g)} ]  \\
 + 2 \bar{\Delta}_{LM}[ \bar{T}_{lm}^{(g)}, \bar{T}_{lm}^{(f)} ] \bar{\Delta}_{LM}[ \bar{T}_{lm}^{(g)}, \bar{T}_{lm}^{(f)} ]
\Bigg>_{g,f}.
\label{eqn:dl}
\end{multline}
For both spectra, the ensemble average is taken over two sets of statistically independent Monte Carlo simulations of $\bar{T}$, labelled $g$ and $f$.
This expression could equivalently be written with just a single set of simulations, however evaluating it as is done here reduces the susceptibility to numerical noise in the evaluation procedure.

Rather than a single mode $C_L^{\Delta \Delta}$, we may be interested in the amplitude $A$ of a fiducial power spectrum $C_L^{\Delta \Delta} = A C_L^{\Delta \Delta, ({\rm fid})}$, for which the corresponding optimal estimator is
\be
\hat{A} = \norm \sum_{L=L_{\rm min}}^{L_{\rm max}} (2L+1) \bar{C}_L^{\Delta \Delta} C_L^{\Delta \Delta, ({\rm fid})},
\label{eqn:hata}
\ee
where $\norm$ is an overall normalization, not to be confused with $\mathcal{N}_{L}$. 
Note the use of the un-normalized $\bar{C}_L^{\Delta \Delta}$ here rather than $\hat{C}_L^{\Delta \Delta}$. 

\subsection{Filtering}
\label{sec:estimator:filtering}
In the case of full-sky coverage with homogeneous noise levels, the inverse-variance filter is given simply by
\begin{align}
\bar{T}_{lm} = \frac{F_{l}}{B_l} \sum_{p=0}^{n_{\rm pix}} \frac{4\pi}{n_{\rm pix}} Y_{lm}(\hat{n}_p) T^{\rm obs}_p,\label{tfil}
\end{align}
where $B_l$ is the beam- and pixel-transfer function, $T^{\rm obs}_p$ is the observed map indexed by pixel $p$ and the filter function $F_l$ is given by
\be
F_l = \frac{1}{C_l^{\rm TT} + C_l^{\rm TT, {\rm noise}}}.
\ee
For the more realistic case of a beam-convolved sky map with inhomogeneous noise, the construction of $\bar{T}_{lm}$ is more involved.
To obtain $\bar{T}_{lm}$ from a set of WMAP sky maps we use an inverse-variance filter which properly accounts for sky-cuts and the inhomogeneity of the map and its noise levels by solving the equation
\begin{multline}
\bar{T}_{lm} = \left( C_l^{\rm TT} \right)^{-1}
\sum_{l' m'} \sum_{p, \nu} 
{\cal C}_{lm, l'm'}^{-1}
{\cal Y}_{l'm'}^{* p, \nu} N_{p, \nu}^{-1} T^{\rm obs}_{p,\nu},
\label{eqn:cinvfilt}
\end{multline}
where the matrix ${\cal C}_{lm, l'm'}$ is given by
\be
{\cal C}_{lm, l'm'} \equiv
\sum_{p,\nu}
\left[
\left( C_l^{\rm TT} \right)^{-1} \delta_{l l'} \delta_{m m'} \\ + {\cal Y}_{lm}^{* p \nu} N_{p, \nu}^{-1} {\cal Y}_{l'm'}^{p, \nu}
\right].
\label{eqn:cinvmat}
\ee
Here the $p$ denotes a map pixel, $\nu$ denotes a particular channel map (usually a given frequency band), $T^{\rm obs}_{p,\nu}$ is the observed (beam convolved) sky map in pixel $p$ at frequency $\nu$, the pointing matrix ${\cal Y}_{lm}^{p, \nu} \equiv B^{\nu}_{l} Y_{lm}(\hat{n}_p)$ gives the value of the spherical harmonic at the center of pixel $p$, convolved with the appropriate beam+pixel transfer function $B_{l}$. $N$ represents is the noise covariance matrix.
We use a diagonal noise covariance in pixel space with
\begin{equation}
N_{p,\nu}^{-1} = \frac{N_{{\rm hits}, p, \nu}}{\sigma_{\nu}^2} M_p,
\label{eqn:ncov}
\end{equation}
where $\sigma_{\nu}$ is a map-dependent noise level and $N_{{\rm hits}, p}$ is the number of observations of pixel $p$ for map $\nu$.
$M_{p}$ is a map which is zero for masked pixels and unity elsewhere. Effectively, it sets the noise level to infinity for masked pixels, ensuring that they are ignored in the rest of the analysis.
There are known noise correlations in the WMAP maps \citep{Jarosik:2010iu} on scales $l < 48$, though we do not incorporate them in our filtering. 
These modes are signal-dominated in the temperature maps which we use, and so this neglect of large-scale noise correlations does not affect our analysis.
We evaluate the matrix inverse of Eq.~\eqref{eqn:cinvmat} using conjugate descent with the fast multigrid preconditioner of Ref. \cite{kendrick_cmb_lens_detect}.
Generally, we will find it useful to work with individual WMAP frequency maps (using only a single entry for $\nu$ in the equations above), however for our final results we will combine all off the useable bands.

For analytical purposes, it is useful to have a diagonal approximation to the full-blown inverse variance filter.
For this we use the $F_l$ function at the top of this section, estimating the average noise power spectrum of the channel-combined, beam-deconvolved map as
\be
C_l^{\rm TT,{\rm noise}} \approx 
\frac{4\pi}{n_{\rm pix} f_{\rm sky}}
\left[
 \sum_{\nu}  \left( B^{\nu}_l \right)^2 
 \frac{1}{\sum_{p=0}^{n_{\rm pix}} N_{p, \nu}}
 \right]^{-1},
\ee
where
$f_{\rm sky} = \sum_{p=0}^{n_{\rm pix}} M_p  / n_{\rm pix}$
is the unmasked sky fraction.

\subsection{Normalization and bias}
\label{sec:normalization}
We now analytically derive an approximate normalization $1/\mathcal{N}$ of our estimator. We will eventually use but correct for deviations from this normalization using Monte Carlo simulations. The analytical treatment, however, is also useful as a tool to explicitly compute the estimator bias induced by other physical sources of non-Gaussianity at the trispectrum level. 

Consider full-sky coverage, with homogeneous noise, in which case the filtered CMB covariance matrix $\bar{C}_{lm, l'm'}$ is diagonal.
If we ensemble average over CMB and CIP realizations in Eq.~\eqref{eqn:triest}, we find that (by construction) $\hat{C}_L^{\Delta \Delta}$ is directly proportional to the connected four point function of $T_{lm}$
\begin{multline}
\langle \hat{C}_L^{\Delta\Delta} \rangle_{{\rm CIP}, {\rm CMB}} =
\frac{\norm_L}{2L+1}
\sum_{M} \sum_{l_i m_i} (-1)^M {\cal G}^{l_1 m_1 l_2 m_2 L}_{l_3 m_3 l_4 m_4 M} 
\\ \times
C_L^{\Delta \Delta, ({\rm fid})}
F_{l_1} F_{l_2} F_{l_3} F_{l_4} 
W^{\Delta}_{l_1 l_2 L} W^{\Delta}_{l_3 l_4 L}  \\ \times
\left<  {T}_{l_1 m_1} \bar{T}_{l_2 m_2} {T}_{l_3 m_3} {T}_{l_4 m_4} \right>_C.
\label{eqn:aresp}
\end{multline}
Here $F_l$ are the filter functions of the previous section.

For generality, so that we can estimate possible biases to our estimator from known non-Gaussian sources such as unresolved point sources and gravitational lensing by large-scale-structure, let us consider the response of this estimator to a trispectrum with the primary form
\be
{}^{x}\! P^{l_1 l_2}_{l_3 l_4} (L) =
C_L^{xx} W^{x}_{l_1 l_2 L} W^{x}_{l_3 l_4 L}.
\ee
This covers both the CIP trispectrum of Eq.~\eqref{eqn:cip_primary_contraction}, as well as the trispectra due to gravitational lensing of the CMB by large-scale-structure, and unresolved point sources which pollute the map.
The trispectrum for CMB lensing is given by  \cite{Hu:2001fa}
\be
{}^{\phi}\! P^{l_1 l_2}_{l_3 l_4} (L) =
C_L^{\phi\phi} W^{\phi}_{l_1 l_2 L} W^{\phi}_{l_3 l_4 L},
\ee
where
\begin{align}
W^{\phi}_{l_{1}l_{2}L} =& \frac{C_{l_{2}}}{2}\sqrt{\frac{\left(2l_{1}+1\right)\left(2l_{2}+1\right)\left(2L+1\right)}{4\pi}}\wigner{l_{1}}{0}{l_{2}}{0}{L}{0}\nonumber \\&\left[L(L+1)+l_{2}(l_{2}+1)-l_{1}\left(l_{1}+1\right)\right]\nonumber\\+&\left\{l_{1}\leftrightarrow l_{2}\right\}.
\end{align}
The trispectrum associated with point source shot-noise is given by \cite{Osborne}
\be
{}^{S^4}\! P^{l_1 l_2}_{l_3 l_4} (L) =
\frac{1}{3} \langle S^4 \rangle W^{S^2}_{l_1 l_2 L} W^{S^2}_{l_3 l_4 L},
\ee
where $\langle S^4 \rangle$ is the kurtosis of the point sources and the weight function is given by
\be
W^{S^2}_{l_{1}l_{2}L}=\wigner{l_1}{0}{l_{2}}{0}{L}{0}\sqrt{\frac{\left(2l_{1}+1\right)\left(2l_{2}+1\right)\left(2L+1\right)}{4\pi}}.
\ee
We discuss effective amplitudes for $S^4$ using the WMAP point source masks in Appendix ~\ref{ps_sec}.

Both point sources and gravitational lensing represent potential sources of bias in the reconstruction of $C_{L}^{\bar{\Delta}\bar{\Delta}}$ and the estimator $\hat{A}$. 
It is important to either verify that these sources of bias are negligible, or to construct appropriately debiased estimators (as done, for example, for estimators of patchy reionization optical depth $\tau$, in Ref. \cite{sudav}, or for CMB lensing estimators in \cite{Namikawa:2012pe}).

We propagate the trispectrum described by the primary contraction through to Eq.~\eqref{eqn:aresp}, obtaining the ensemble-averaged contribution of the physical effect $x$ (which can denote CIPs, weak gravitational lensing, or point sources) to the CIP trispectrum estimator $\hat{A}$,
\be
\left. \left< \bar{C}_L^{\Delta\Delta} \right> \right|_{x} = 
{\cal P}_L^{\Delta x} + {\cal S}_L^{\Delta x}
.\label{hatexp}
\ee
Here ${\cal P}_L^{\Delta x}$ and ${\cal S}_L^{\Delta x}$ capture the contributions from the primary and secondary contractions of any trispectrum respectively.

The primary term is given by
${\cal P}_L^{\Delta \Delta} = C_L^{xx} {\cal R}_L^{\Delta x}$, where the response function ${\cal R}_L^{\Delta x}$ is given by
\be
{\cal R}_L^{\Delta x} = \left[ \frac{1}{2L+1} \sum_{l l'} W_{l l' L}^{\Delta} W_{l l' L}^{x} F_{l} F_{l'} \right]^2.
\label{eqn:resp}
\ee
The secondary term ${\cal S}$ is more complicated, involving Wigner-$6$j symbols which are numerically intensive to calculate.
We estimate these contributions using flat-sky expressions to evaluate the secondary contractions, given by \cite{n1bias}
\begin{multline}
{\cal S}_L^{\Delta x} = 
\int \frac{d^2 l_1}{(2\pi)^2} 
\int \frac{d^2 l_2}{(2\pi)^2} 
F_{ | \bl_1 |} F_{ | \bl_2 | }
W^{\Delta}( \bl_1,  \bl_2)
W^{\Delta}( \bl_1', \bl_2')
\\ \times
\Big\{
C_{| \bl_1 - \bl_1' |}  
W^{x}( -\bl_1,  \bl_1' )
W^{x}( -\bl_2, \bl_2' ) \\
+
C_{| \bl_1 - \bl_2' |}  
W^{x}( -\bl_1,  \bl_2' )
W^{x}( -\bl_2, \bl_1' )
\Big\}\label{bigflat},
\end{multline}
where $\bl_{1}$ and $\bl_{2}$ are Fourier space angular multipole vectors, $\bl_1 + \bl_2 = \mathbf{L}$ and the flat-sky weight function is given by
\be
W^{\Delta}(\bl_1, \bl_2) = C_{|\bl_1|}^{{\rm T, dT}} + C_{|\bl_2|}^{{\rm T, dT}}.
\ee
The flat-sky weight functions for CMB lensing and point source shot noise are \cite{2002ApJ...574..566H,Osborne}
\begin{align}
W^{\phi}(\bl_1,\bl_2)&=C_{l_{1}}^{\rm TT}\left[\left(\bl_{1}+\bl_{2}\right)\cdot \bl_{1}\right] + (\bl_1 \leftrightarrow \bl_2), \nonumber \\
W^{S^2}(\bl_1,\bl_2)&=1.
\end{align}

A special case of the bias calculation is for $x=\Delta$, which yields the response of $\bar{C}_L^{\Delta \Delta}$ to CIP fluctuations themselves and therefore the normalization of the estimator.
We will see that for CIPs, the primary term is dominant. 
We can therefore use Eq.~\eqref{eqn:resp} as an approximate, analytical normalization for $\hat{C}_L^{\Delta \Delta}$, 
with
\be
\norm_L^{\textsc{approx}} = \left[ f_{\rm sky} {\cal R}_L^{\Delta \Delta} \right]^{-1}
\label{eqn:normlapprox}
\ee
The corresponding approximate normalization for the fiducial power spectrum amplitude estimator $\hat{A}$ is given by
\be
\norm^{\textsc{approx}} = \left[ f_{\rm sky} \sum_{L=L_{\rm min}}^{L_{\rm max}} (2L+1) \left( C_L^{\Delta \Delta, ({\rm fid})} \right)^2 {\cal R}_L^{\Delta \Delta} \right]^{-1}.\label{napprox}
\ee
We will ultimately correct this normalization for our $\hat{A}$ estimates using Monte Carlo simulations which implicitly include the contribution from the secondary contractions above, as well as cut-sky effects beyond simple $f_{\rm sky}$ scaling. For a scale-invariant power spectrum we find that the approximation of Eq.~\eqref{napprox} is accurate to better than $20\%$.

\begin{figure}[htbp]
\includegraphics[width=\columnwidth]{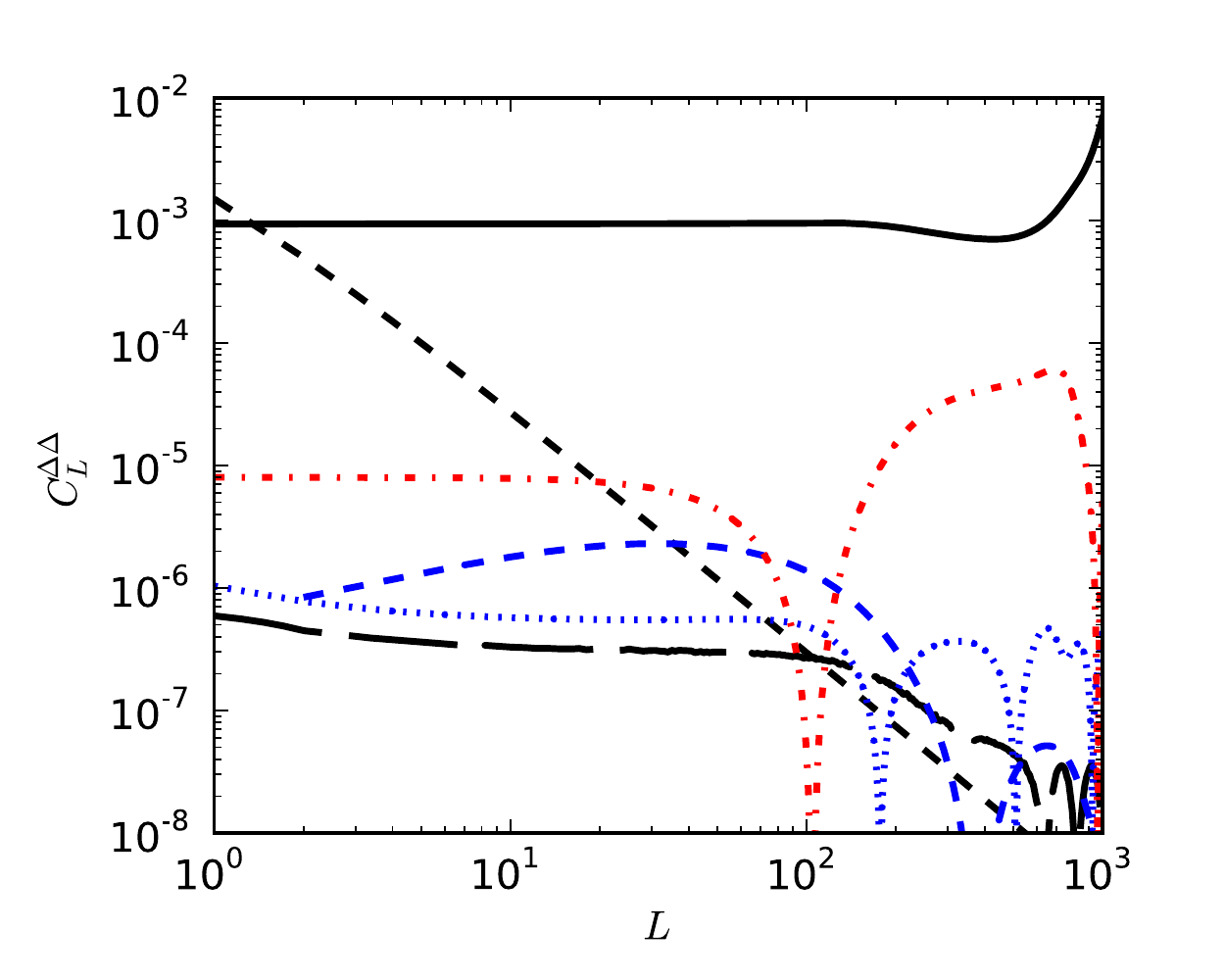}
\caption{
Contributions to our trispectrum estimator of the normalized CIP power spectrum $C_{L}^{\Delta \Delta}$ as a function of scale, for WMAP V-band noise and beam.
The estimator noise power spectrum ${\cal N}_L^{\textsc{approx}} D_L^{\Delta\Delta}$ is shown as black solid.
For comparison, we plot a scale invariant power spectrum $C_L^{\Delta \Delta} = 0.003 \times [L(L+1)]^{-1}$ with an amplitude comparable to our one standard-deviation constraint (thick black dashed). 
We also calculate the contribution from the secondary contractions for this CIP power spectrum (${\cal N}_L^{\textsc{approx}} {\cal S}_L^{\Delta \Delta}$, black long-dashed).
The expected contribution from the primary and secondary contractions of the CMB lensing trispectrum are shown as blue dashed and blue dotted respectively.
In addition, we plot a pessimistic (larger than the maximum in Appendix~\ref{ps_sec}) estimate of the V-band contamination from unresolved point sources, with $S^4 = 5\times10^{-7} \mu K^4$ (red dash-dotted).
The primary and secondary contributions from the shot-noise trispectrum have the same shape, and so we only plot their sum. An analogous analysis in the Q and W-bands yields the same conclusion, and so we omit those curves here.}
\label{fig:cldd_sources}
\end{figure}

In Fig.~\ref{fig:cldd_sources} we show estimates of the contribution of primary and secondary contractions of CIPs for a scale-invariant power spectrum, as well as weak lensing of the CMB and point sources.
Using Fig.~\ref{fig:cldd_sources}, we see that the primary and secondary contractions of the lensing trispectrum yield negligible contributions to the CIP estimator, compared with the estimator noise power spectrum. We also see that the point source (discussed at length in Appendix \ref{ps_sec}) trispectrum contributes negligibly to the CIP estimator.
We also see that secondary contractions of the CIP trispectrum are negligible compared with primary contractions.

\label{bias}
\section{WMAP Analysis}
\label{bam}

\subsection{Data and Simulations}
We analyze the ``foreground-reduced'' CMB temperature anisotropy maps, with associated beam and noise characterization, from the final, 9-year WMAP data release \cite{Bennett:2012fp}.
We use the Q, V, and W-band maps, with nominal central frequencies of 40, 60, and 90GHz respectively.
To remove known Galactic and bright point source contamination we use the ``KQ85'' temperature analysis mask produced by the WMAP team.

To compute the mean-field and disconnected noise bias of our estimator [Eqs.~(\ref{eqn:dl}) and (\ref{eqn:cl})], we use CMB+noise simulations of the data to compute the Monte Carlo averages ($300$ realizations per WMAP band). We simulate a CMB temperature realization $T_{lm}^{\rm sim}$ with power spectrum $C_l^{\rm TT}$ given by a flat $\Lambda$CDM cosmology consistent with the WMAP power spectrum \cite{Hinshaw:2012fq}, with parameters given in the introduction, computed using the \textsc{camb} \cite{camb} code.
We apply a beam and pixel transfer function to these temperature multipoles, and then project them onto an $N_{\rm side}=512$ HEALPix map with a harmonic transform. 
To each pixel in this map, we add a Gaussian noise contribution with variance given by Eq.~\eqref{eqn:ncov}.
These simple simulations are then filtered in the same way as the real data [that is, using Eq.~(\ref{tfil})] to produce the simulated quantities $\bar{T}^{(g)}$ and $\bar{T}^{(f)}$ of Eqs.~\eqref{eqn:dl} and \eqref{eqn:cl}.

To test the normalization of our CIP estimator, we also require non-Gaussian simulations with $C_L^{\Delta \Delta} \ne 0$.
We form these non-Gaussian CMB simulations with power spectrum $C_l^{\rm TT}$ and trispectrum given by Eq.~\eqref{eqn:cip_primary_contraction} using three pieces $T_{lm} = A_{lm} + \beta_{lm} + C_{lm}$.
$A_{lm}$ is drawn from a Gaussian distribution with power spectrum $C_l^{\rm AA}$ (the choice of $C_l^{\rm AA}$ is somewhat arbitrary for our purposes, we will use $C_l^{\rm AA} = C_l^{\rm TT}/2)$. 
Then we add the term $\beta_{lm}$ which ensures that our simulations have the desired trispectrum, with
\begin{multline}
\beta_{lm}
=
\int d\hatn Y_{lm}^*(\hatn) 
\left[ \sum_{LM} Y_{LM}(\hatn) \Delta_{LM} \right]  \\ \times
\left[ \sum_{l'm'} Y_{l'm'}(\hatn) A_{l'm'} \frac{C_{l'}^{\rm T, dT}}{C_{l'}^{\rm AA}} \right].
\end{multline}
Here $\Delta_{LM}$ are Gaussians drawn for a fiducial power spectrum $C_L^{\Delta \Delta, ({\rm fid})}$.
We add a final Gaussian term $C_{lm}$ which sets the overall power spectrum of our non-Gaussian simulations to be $C_l^{\rm TT}$. 
The $C_{lm}$ are drawn from the power spectrum $C_l^{CC} = C_l^{\rm TT} - C_l^{\rm AA} - C_l^{\beta\beta}$, with
\begin{multline}
C_l^{\beta\beta} = \sum_{l l'} \frac{(2l+1)(2l'+1)}{4\pi} \threej{l}{l'}{L}{0}{0}{0}^2  \\ \times
\frac{ C_L^{\Delta \Delta} \left(C_{l'}^{\rm T, dT} \right)^2}{C_{l'}^{\rm AA}} .
\end{multline}

\begin{figure*}[h]
\includegraphics[width=\textwidth]{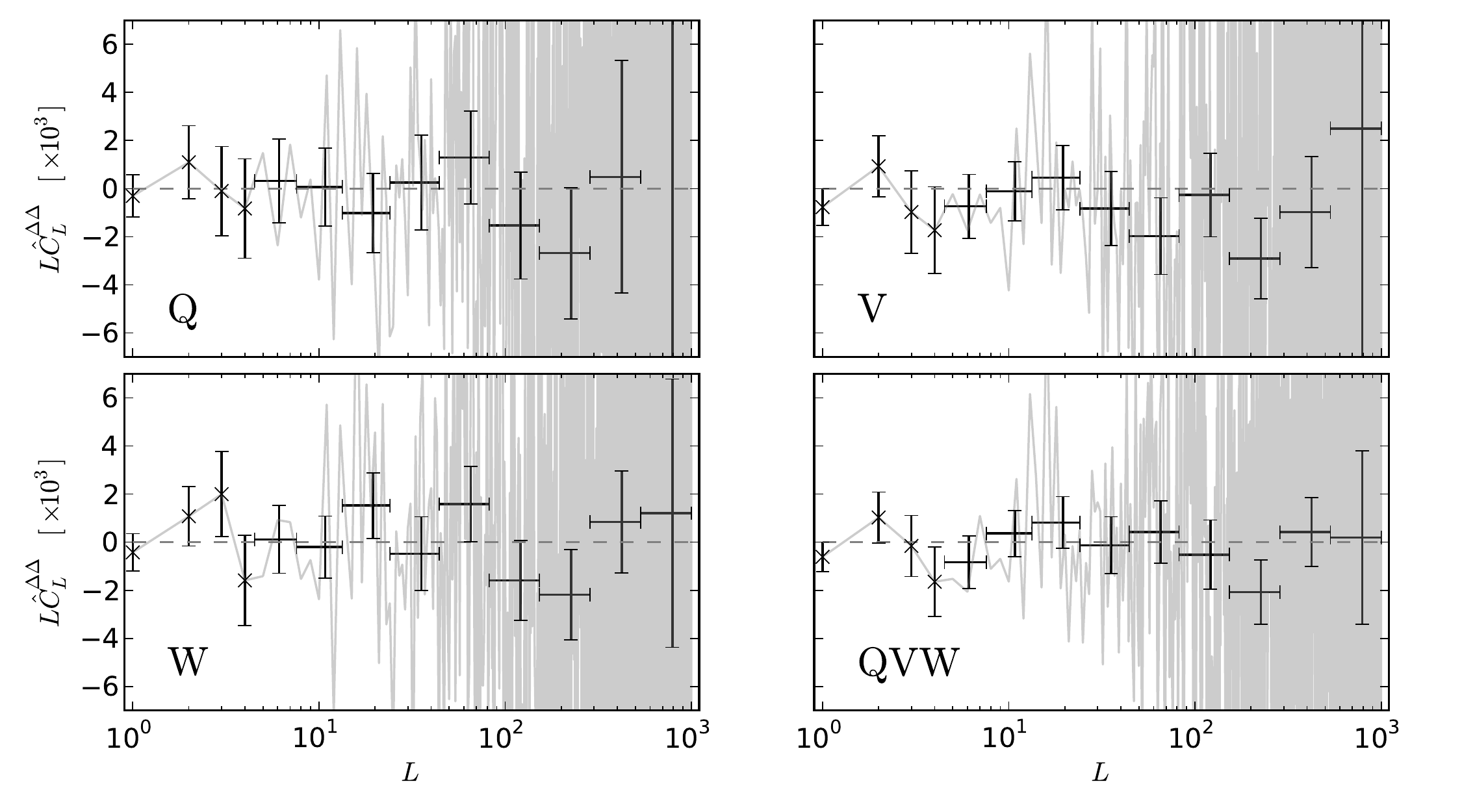}
\caption{Estimates of the CIP power spectrum $C_L^{\Delta \Delta}$ using the WMAP Q, V, and W-band data. The ``QVW'' points give the result when all three bands are combined with a inverse-variance filter. Gray lines give the measurements for individual multipoles.}
\label{fig:cldd_specs}
\end{figure*}

\begin{figure*}[h]
\includegraphics[width=4.0 in]{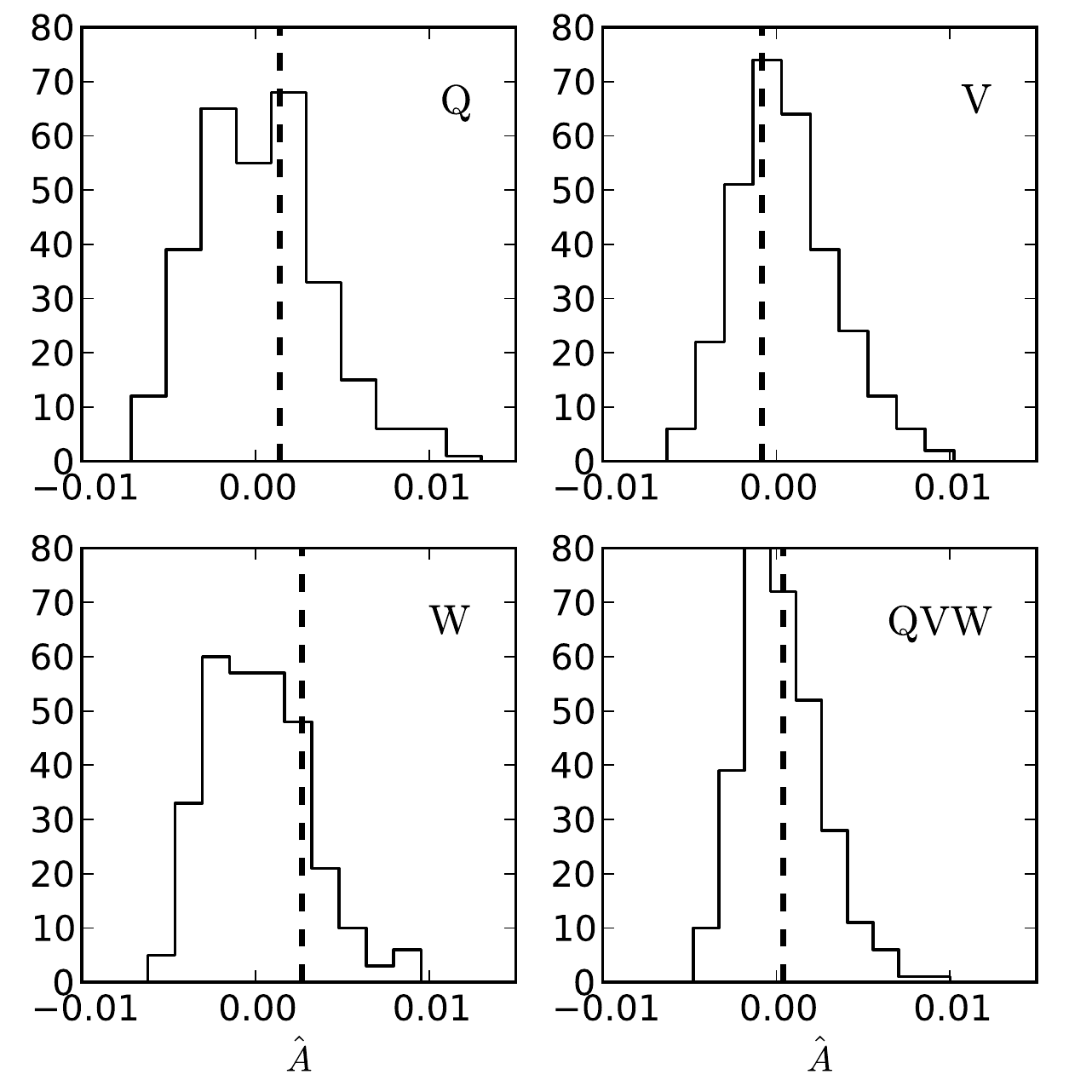}
\caption{
Fits for the amplitude of a scale invariant power spectrum [Eq.~(\ref{eqn:scalinv})], using the WMAP Q, V, and W-band data. 
Dashed vertical lines give the fits themselves, while the solid histograms give the distribution of $\hat{A}$ measured on simulations with $\hat{A}=0$.}
\label{fig:cldd_amp_hist}
\end{figure*}

\subsection{Results}
\label{rsec}
In Fig.~\ref{fig:cldd_specs} we plot estimated CIP power spectra $\hat{C}_L^{\Delta \Delta}$ from separate analyses of Q, V, and W-band maps, as well as an inverse-variance weighted combination of the $3$ maps (denoted QVW). We use the approximate analytical normalization of Eq.~\eqref{eqn:normlapprox}. This
neglects the secondary CIP trispectrum contractions, which we see in Fig.~\ref{fig:cldd_sources} are negligible, as well as cut-sky effects beyond simple $f_{\rm sky}$ scaling. For the question we are asking in this figure (``Is the measured CIP power non-zero with statistical significance?'') the accuracy of the normalization is not important, as corrections to Eq.~\eqref{eqn:normlapprox} show up in both the signal and noise, and thus leave the significance of any signal or limit unaffected. We see no evidence for anomalous $C_L^{\Delta \Delta}$ power in the individual Q, V, or W-band maps. We may now obtain limits to specific models for the CIP power spectrum, as well as model-independent constraints to the CIP power spectrum at different multipole numbers $L$.
\subsubsection{Limits to a scale-invariant power spectrum of CIPs}
We now place limits to the amplitude of a scale-invariant power spectrum of CIP fluctuations, which in multipole space projected onto a two-dimensional sky is \cite{grin_prd}
\be
C_L^{\Delta \Delta} =C_{L}^{\Delta \Delta,(\rm fid)} \simeq \frac{A}{L(L+1)},
\label{eqn:scalinv}
\ee
for some amplitude $A$. For $A=1$, corresponding realizations of $\Delta$ would have fluctuations with amplitude of approximately unity. For $L\gsim 870$, the scaling in Eq.~(\ref{eqn:scalinv}) actually turns over to $L^{-3}$ as the wavelength of the CIP falls below the thickness of the last-scattering surface. These modes have very low signal-to-noise ratio using present data and thus contribute negligibly to our estimator of $A$. It is thus adequate to use Eq.~(\ref{eqn:scalinv}). The results of these fits are shown in Fig.~\ref{fig:cldd_amp_hist}, and are also consistent with zero. Note that a value of $A < 0$ is non-physical, but allowed by our measurement as we have not imposed a prior on the positivity of $A$.

For our measurement, we use Eq.~\eqref{eqn:hata}, $L_{\rm min} = 2$, and $L_{\rm max}=1000$. We use the normalization
${\cal N} = {\cal N}^{\textsc{approx}}$,
where ${\cal N}^{\textsc{approx}}$ is given in Eq.~\eqref{napprox}. Although this normalization is only approximate, it is adequate to answer the question of whether or not the measured value of $A$ deviates from zero with statistical significance. Later, when limits to $A$ are set, we account for corrections to this normalization.

Our constraint on $\hat{A}$ comes mainly from very low-$L$ modes of $\bar{C}_L^{\Delta \Delta}$. The noise variance of $\hat{C}_L^{\Delta \Delta}$ at $L<100$ goes as $1/(2L+1)$, which falls off more slowly than the scale-invariant spectrum, and so most of our sensitivity to ${A}$ comes from a small number of low-$L$ modes.  Our estimator probability density function (PDF) is thus slightly non-Gaussian (as can be seen from the histograms of Fig.~\ref{fig:cldd_amp_hist}), for the same reason that the distribution of low-$L$ power-spectrum estimates is non-Gaussian. This also occurs for trispectrum-based estimators of the amplitude of local-type non-Gaussianity $|f_{\rm NL}|$ \cite{Smith:2012ty}. 

We find best-fit values in Q, V, and W-bands separately, as well as for the QVW combination, of $\hat{A}_{\rm Q}\simeq 1.4 \times 10^{-3}$, $\hat{A}_{\rm V}\simeq -8.4 \times 10^{-4}$, $\hat{A}_{\rm W}\simeq2.7\times 10^{-3}$, $\hat{A}_{\rm QVW}\simeq 3.9\times 10^{-4}$. In Fig.~\ref{fig:cldd_amp_hist}, these values are compared with the distribution of $\hat{A}$ values for simulations in which the real value $A=0$. We see that our best-fit values are consistent with the null hypothesis at $95\%$-confidence.

To establish upper limits to $A$, we need the function $P(A|\hat{A})$, the probability distribution of $A$ values, given the estimated value $\hat{A}$. Assuming flat priors for the `data' ($\hat{A}$) and theoretical parameter ($A$), we have $P(A|\hat{A})\propto P(\hat{A}|A)$. From our Monte Carlo simulation of the $A=0.01$ case, we know that the width of the non-Gaussian function $P(\hat{A}|A)$ depends noticeably on the real value of $A$, and so it is important to properly determine $P(\hat{A}|A)$.

\begin{figure*}[h]
\includegraphics[width=6.0 in]{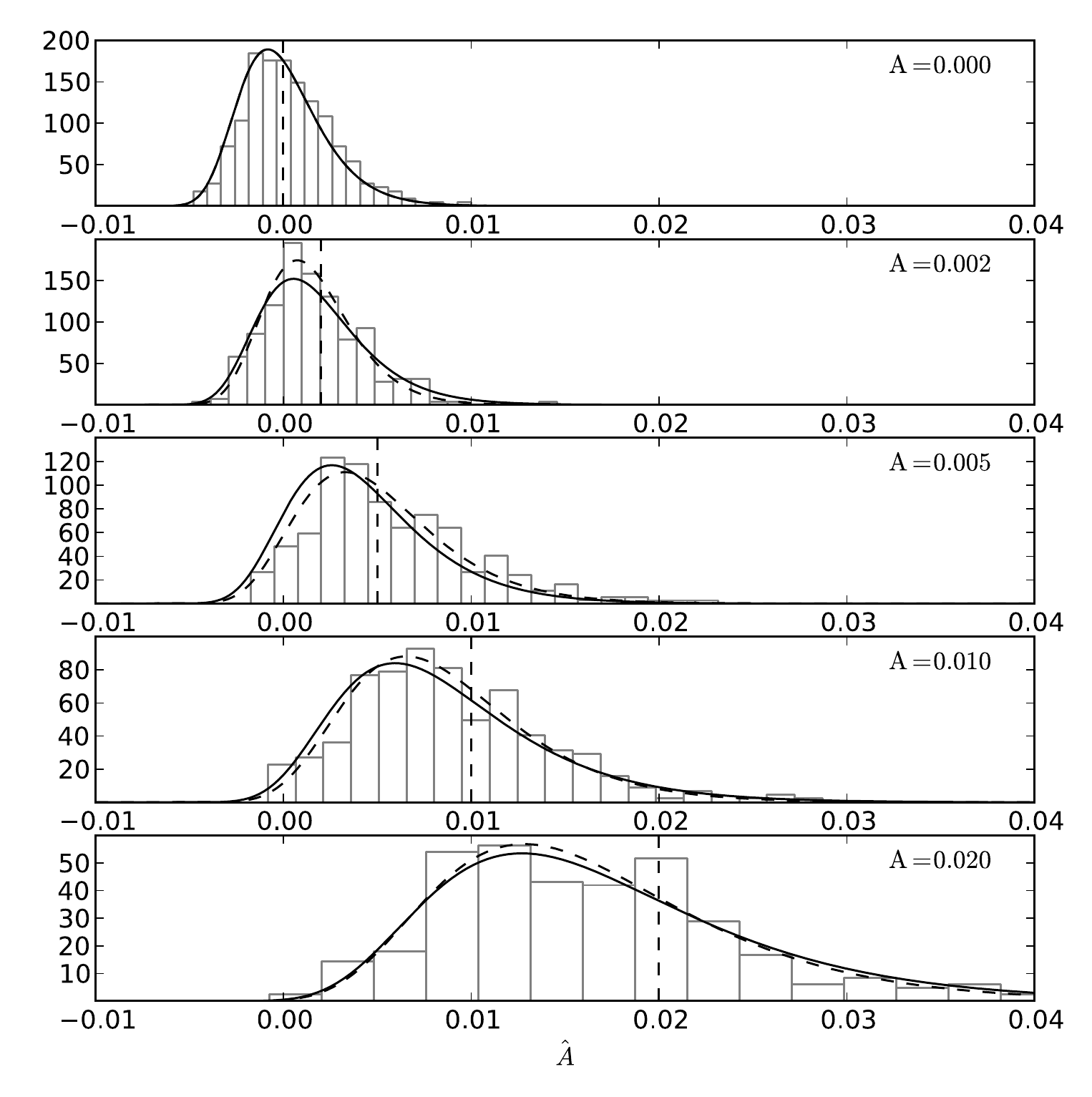}
\caption{
Arbitrarily normalized probability distribution for $\hat{A}_{\rm QVW}$ in the presence of a scale-invariant spectrum of CIPs $C_{L}^{\Delta \Delta}=A/\left[L(L+1)\right]$. The estimated value is determined from an appropriately inverse-variance weighted sum of the $3$ maps, simulated with an ensemble of Monte Carlo simulations. Dashed vertical line shows input value of $A$. Dashed curves show $P^{\alpha}(\hat{A}|A)$, while solid curves show $P^{\beta}(\hat{A}|A)$. Both models are described in Sec. \ref{rsec}.}
\label{fig:cip_pdf}
\end{figure*}
To this end, we conduct a suite of Monte Carlo simulations, estimating $\hat{A}_{\rm QVW}$ for multiple realizations ($300$ per band and per real $A$ value), at the series of true parameter values $A=\left\{0.000,0.002,0.005,0.010,0.020\right\}$. The results for $P(\hat{A}|A)$ are shown in Fig. \ref{fig:cip_pdf}. To compute an upper limit to $A$, we use a model for $P(A|\hat{A})$ [and thus $P(\hat{A}|A)$]. 

The model for $P(\hat{A}|A)$ is as follows: A $\chi^{2}$ distribution $P_{\chi}(\mathcal{E}_{L}; k_{L}, s_{L})$ is fit to the simulated random variable $\mathcal{E}_{L}\equiv \mathcal{N}\left(2L+1\right)\overline{C}_{L}^{\Delta \Delta}C_{L}^{\Delta \Delta,{\rm fid}}$ for $2\leq L\leq 20$. At each $L$ and for any given $A$ value, this distribution has two free parameters, the number of degrees of freedom $k_{L}$, and a scale parameter $s_{L}$ such that $\left \langle \mathcal{E}_{L}\right \rangle=k_{L}s_{L}$. The estimator used is given by \be\hat{A}=\sum_{L=L_{\rm min}}^{L=L_{\rm max}}\mathcal{E}_{L}.\ee A model PDF $P^{\alpha}(\hat{A}|A)$ is thus obtained by convolution over $P_{\chi}(\mathcal{E}_{L}; k_{L}, s_{L})$. This model is shown as the dashed curve in Fig.~\ref{fig:cip_pdf} and is clearly a reasonable fit to the data. 

Alternatively, we may just fit $P_{\chi}(\mathcal{E}_{L}; k_{L}, s_{L})$  for the $A=0$ case and then adjust the scale parameter $s_{L}\to s_{L}\times  \left\{(1+A/\left[L\left(L+1\right)\left(2L+1\right)\right]\right\}$. An analogous convolution then yields the semi-analytic function $P^{\beta}(\hat{A}|A)$, which is comparably accurate to $P^{\alpha}(\hat{A}|A)$ in fitting the estimator PDF from the simulation, and readily interpolated to obtain quantities of interest. 

Our final upper limit is computed by starting with $P^{\beta}(\hat{A}|A)$, obtaining $P(A|\hat{A})$ using Bayes's theorem with a flat prior on $A$ in the simulated domain, computing the cumulative probability function (CDF) associated with $P(A|\hat{A})$, and interpolating this function, incrementing $A$ upwards to define a $95\%$-confidence interval. Like the measurement of $\hat{A}_{\rm QVW}$, these simulations use $\mathcal{N}=\mathcal{N}^{\rm approx}$, and thus calibrate the relationship between $A$ and $\hat{A}_{\rm QVW}$, accounting for errors induced in the approximation $\mathcal{N}=\mathcal{N}^{\rm APPROX}$. The resulting upper limits to $A$ are thus correctly normalized.

Given the observed multi-band value $\hat{A}_{\rm QVW}=3.9\times 10^{-4}$, we compute $P(3.9\times 10^{-4}|A)$ [and thus $P(A|3.9\times 10^{-4})$], and find that $A\leq 1.1\times 10^{-2} $ at $95\%$ confidence. Measurements of the baryon fraction in galaxy clusters \cite{holder} impose the limit $A\leq 5.4\times 10^{-3}$ to the amplitude of a scale-invariant spectrum of CIPs \cite{grin_prl,grin_prd}. The larger value quoted in Ref. \cite{grin_prl} is obtained using a different definition of $A$, which we have appropriately rescaled here. We may also restate our result in terms of $\Delta_{\rm cl}$, the RMS fluctuation in the baryon-dark-matter density ratio on galaxy cluster scales, using our definition of $A$ and Eq.~(1) of Ref. \cite{grin_prl} (with the associated wave-number range $k_{\rm min}\lesssim k\lesssim k_{\rm max}$) to obtain
\begin{equation}\Delta_{\mathrm{cl}}^2\simeq \frac{A
\ln(1000)}{2\pi}.\end{equation}
We obtain $\Delta_{\rm cl}\lesssim 0.11$, to be compared with the result directly obtained from the cluster baryon fraction, $\Delta_{\rm cl}\lesssim 0.077$.

As predicted in the forecasts of Ref. \cite{grin_prd}, the CIP-sensitivity of WMAP is comparable to measurements of the baryon fraction in galaxy clusters. The physics of this CMB probe of CIPs, however, is completely different than that used in the cluster probe, and offers an important and truly primordial constraint on the amplitude of CIPs. 
\label{limsec}

\subsubsection{Constraints to model-independent CIP amplitude at different angular scales}
Aside from placing limits to a scale-invariant CIP power spectra, we may also place limits to the CIP amplitude at different angular multipole numbers $L$, without reference to a fiducial model.
\begin{table}[htbp]
\begin{center}
\begin{tabular}{ | c | c | c | c  | p{7cm} |}
\hline
L & $C_L^{\Delta \Delta,\rm max}$& $\theta$ (in $^{\circ}$)&$\Delta_{\rm rms}^{\rm max}$ \\ \hline \hline
1 & $2.7\times 10^{-2}$& 100 & $9.2 \times 10^{-2}$  \\ \hline
2 & $2.1\times 10^{-2}$& 50 & $1.4\times 10^{-1}$\\ \hline
3 & $4.3\times 10^{-3}$& 33 & $9.1\times 10^{-2}$\\ \hline
4 & $1.7\times 10^{-3}$ & 25&$7.3\times 10^{-2}$\\ \hline
5 & $1.6\times 10^{-3}$& 20 & $8.8\times 10^{-2}$\\ \hline
6 &$1.0\times 10^{-3}$& 17& $8.3\times 10^{-2}$\\ \hline
7 & $2.0\times 10^{-3}$& 14& $1.3\times 10^{-1}$\\ \hline
8 &$1.0\times 10^{-3} $& 13 & $1.1\times 10^{-1}$ \\ \hline
9 & $9.0\times 10^{-4}$ & 11&$1.1\times 10^{-1}$\\ \hline
10 &$7.8\times 10^{-4} $& 10&$1.2\times 10^{-1}$\\ \hline
11 & $1.2\times 10^{-3}$& 9.1 & $1.6\times 10^{-1}$\\ \hline
12 & $4.6\times 10^{-4}$& 8.3& $1.1\times 10^{-1}$\\ \hline
13 &$9.7\times 10^{-4}$& 7.7 & $1.7\times 10^{-1}$\\ \hline
14 & $7.8\times 10^{-4}$ & 7.1&$1.7\times 10^{-1}$\\ \hline
15 &$4.5\times 10^{-4}$& 6.7 & $1.3\times 10^{-1}$\\ \hline
16 & $6.9\times 10^{-4}$& 6.3 & $1.7\times 10^{-1}$\\ \hline
17 & $4.9\times 10^{-4}$& 5.9 & $1.5\times 10^{-1}$\\ \hline
18 &$5.3\times 10^{-4}$& 5.6 & $1.7\times 10^{-1}$\\ \hline
19 & $3.9\times 10^{-4}$& 5.3 & $1.5\times 10^{-1}$\\ \hline
20 & $4.0\times 10^{-4}$& 5.0 & $1.6\times 10^{-1}$\\ \hline
\end{tabular}
\caption{Upper limits ($95\%$-confidence level) $C_L^{\Delta \Delta,\rm max}$ to the angular power spectrum of CIPs as a function of multipole number $L$, as recovered from an inverse-variance weighted sum of the $3$ WMAP 9-year CMB maps, and the Monte Carlo simulations described in Sec. \ref{limsec}. Also shown is the corresponding angular scale in degrees of $\theta^{\circ}\simeq 100^{\circ}/L$ and the $95\%$-confidence upper limit to the RMS CIP amplitude on that angular scale, $\Delta_{\rm rms}^{\rm max}\simeq \sqrt{L(L+1)C_{L}^{\Delta\Delta,\rm max}/(2\pi)}$.}
\label{tb:cldd_all}
\end{center}
\end{table}
\begin{figure}[htbp]
\includegraphics[width=3.50in]{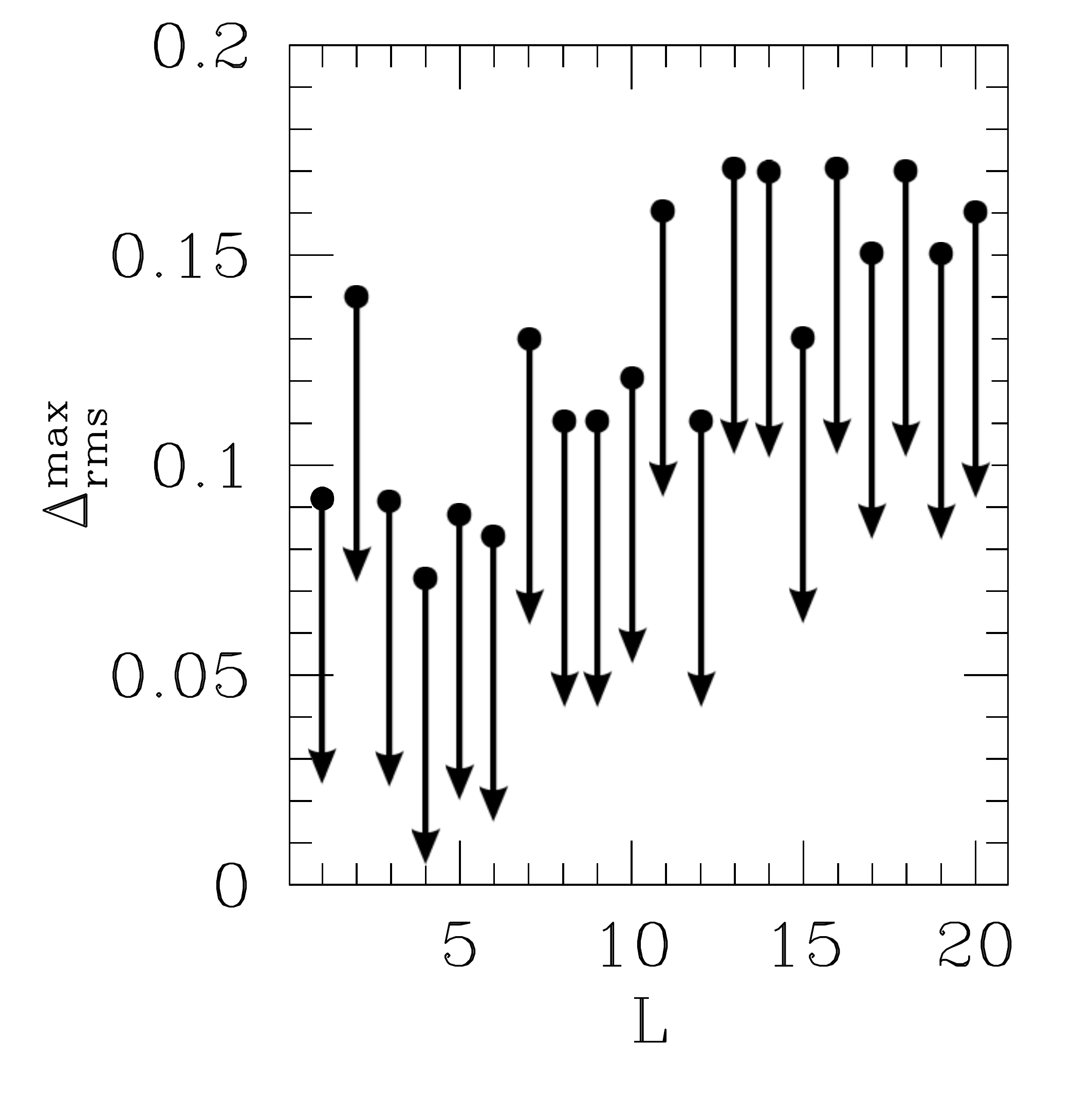}
\caption{Upper limits ($95\%$-confidence level) to the RMS CIP amplitude as a function of multipole number $L$, as recovered from an inverse-variance weighted sum of the $3$ WMAP 9-year CMB maps, and the Monte Carlo simulations described in Sec. \ref{limsec}.}
\label{flimfig}
\end{figure}

The PDF for $P^{\beta}(\hat{A}|A)$ used to obtain the above upper limits is derived by fitting a $\chi^{2}$ distribution to the PDF for each $\hat{C}_{L}^{\Delta \Delta}$, assuming no covariance between $\hat{C}_{L}^{\Delta \Delta}$ for different $L$, convolving over different multipole moments, and applying Bayes's theorem. This \textit{ansatz} fits the Monte Carlo histograms well, and so we may use the same simulation to impose limits to $C_{L}^{\Delta \Delta}$ independently for each scale $L$, without making any assumptions about the functional form of the CIP power spectrum. 

We thus model the histogram of $\hat{C}_{L}^{\Delta \Delta}$ values using the function $P_{\chi}(\mathcal{E}_{L};k_{L},s_{L})$, where $s_{L}=s_{L}^{A=0}\times \left\{[1+C_{L}^{\Delta \Delta}/\left(2L+1\right)]\right\}$. Here $s_{L}^{A=0}$ is the best-fit value of the $\chi^{2}$ scale-parameter $s_{L}$ for the null hypothesis simulation. For all multipoles in the range, $1\leq L\leq 20$, this PDF fits the Monte Carlo results. Applying Bayes's theorem on an $L$-by-$L$ basis with the resulting PDF, we obtain $P(C_{L}^{\Delta \Delta}|\hat{C}_{L}^{\Delta \Delta})$. Using the estimated values $\hat{C}_{L}^{\Delta \Delta}$ from the minimum-variance-weighted QVW combination, we obtain $95\%$-confidence upper limits to $C_{L}^{\Delta \Delta}$ in the $L$-range of interest. The results are shown in Table I and Fig. \ref{flimfig}, where the corresponding angular scale and RMS CIP amplitude upper limit is shown. We thus see that the relative mass fractions of baryons and dark matter can vary by no more than $\sim 10\%$ on angular scales of $5-100^{\circ}$ at the SLS. This limit is forced upon us by CMB data alone, with no reliance on galactic abundance measurements, galaxy physics, or knowledge about the nature of cosmic reionization.

\section{Conclusions}
Compensated isocurvature perturbations (CIPs) break the usual assumption that the primordial cosmic baryon fraction is spatially homogeneous. Here, building on past work searching for CIPs using variations in the cluster baryon fraction, we use WMAP 9-year maps of the CMB temperature fluctuations to search for CIPs. We apply a trispectrum-based estimator to show that the amplitude of a scale-invariant spectrum of CIPs must obey the constraint $A\lsim 1.1\times 10^{-2}$, combining all $3$ CMB-dominated WMAP bands. Independent of the functional form of the CIP power spectrum, the rms CIP amplitude must be no greater than $\sim 10\%$ on scales of $5-100^{\circ}$ when Thomson scattering freezes out at $z\sim 1100$.

As we saw, the estimator for $C_{L}^{\Delta \Delta}$ is an appropriately filtered $4$-pt function of the data, probing correlations that are absent under the null hypothesis. Gaussian CIPs uncorrelated with the underlying adiabatic fluctuation can then be thought of as inducing a non-Gaussian signal in the CMB, with vanishing $3$-pt function and non-vanishing $4$-pt function. 

It may be that CIPs are not an independent Gaussian random field (as assumed here), but rather, as in some curvaton models \cite{gordon_pritchard}, correlated with the usual adiabatic fluctuations. 
In this case, schematically, the induced temperature fluctuation is $\delta T\sim \Delta \Phi$, as before, but now, since $\Delta=A\Phi$ for some correlation coefficient, $\delta T\sim \Phi^{2}$. The resulting CMB $3$-pt function is then $\left \langle T T T\right \rangle \propto \left \langle \delta T ~TT \right \rangle \propto \left \langle \Phi^{4}\right \rangle$. Curvaton-inspired CIPs will thus cause an additional non-zero non-Gaussianity, a $3$-pt correlation or bispectrum. 

We note that the CIP estimator described here could also be applied to the recently released nominal mission data of the \textit{Planck} satellite, for which we forecast a sensitivity improvement factor of $\sim 3$ in $\Delta_{\rm cl}$ or $A$ \cite{grin_prd}. Future ground-based CMB polarization data (such as ACTPol \cite{actpol} and SPTpol \cite{sptpol}) also have the potential to significantly improve the sensitivity of the CIP probe described here.

It is claimed in Ref. \cite{gordon_pritchard} that CIPs would lead to negligible changes in the total matter power spectrum, and would thus not be detectable using galaxy surveys. It stands to reason, however, that star formation efficiency, cooling, and other processes are strongly dependent on the baryon density, and that the statistics of actual galaxies would be drastically altered in the presence of CIPs. Using semi-analytic models of galaxy formation, this could be tested. Large fluctuations in the Jeans/cooling masses in the presence of a CIP could also change the dynamics of reionization, leading to an additional probe. We will explore these possibilities in future work.

 \label{conclusions}


\begin{acknowledgments}
We acknowledge useful conversations with 
D.~N.~Spergel, C.~Dvorkin, and K.~M. Smith. DG was supported at the Institute for Advanced Study by the National Science Foundation (AST-0807044) and NASA (NNX11AF29G). MK was supported by the Department of Energy (DoE SC-0008108) and NASA (NNX12AE86G). 
Part of the research described in this paper was carried out at the Jet Propulsion
Laboratory, California Institute of Technology, under a contract with the National Aeronautics and Space Administration. This work was supported by a  CITA National Fellowship at McGill. Some of the results in this paper have been derived using HEALPix \cite{healpix}.
This work was begun during the 2011 Winter conference ``Inflationary theory and its confrontation with data in the Planck era'' at the Aspen Center for Physics (NSF Grant 1066293). 
The authors are very grateful for the hospitality of the Aspen Center.
\end{acknowledgments}

\appendix
\section{Point Sources}
\label{ps_sec}
Points sources which are bright enough to be detected are removed using the WMAP analysis mask used in this paper. There is, however, a population of residual sources which are too faint to be directly detected individually the maps, but which contribute non-negligibly to the CMB temperature power spectra, as well as higher order statistics like the trispectrum which we probe here.
When modelling these sources, it is useful to consider an effective limiting flux threshold $S_{c}$, above which all sources are assumed to be detected, and below which none are.
At WMAP frequencies and sensitivity, the dominant residual contribution is from a small number of radio sources directly below the detection threshold. 
The statistics of this population are dominanted by the 1-point (or ``shot-noise'') component.
If the differential number density of sources with flux $S$ is denoted $dN/dS$, the unresolved shot noise contribution to the reduced trispectrum is given by \cite{tofu_paper,acoustic_sig,asantha_ps}
\begin{equation}
\left \langle S^{4}\right \rangle \equiv g^{4}(x)\int_0^{S_{c}} dS S^{4} \frac{dN}{dS},\label{trispec_ps}
\end{equation}
where 
$x=h\nu/(k_{\rm B}T)=\nu/56.84~{\rm GHz}$ for experimental channel frequency $\nu$
and $g(x)$ is a conversion factor from Jansky flux units to thermodynamic $\mu {\rm K}$, given by
\begin{eqnarray}
g(x)&\equiv& 2 \frac{\left(hc\right)^{2}}{\left(k_{B}T\right)^{3}} \left[\frac{\sinh^{2}{\left(x/2\right)}}{x^{4}}\right]\nonumber \\&\simeq&\frac{\left(e^{x}-1\right)^{2}}{x^{4}e^{x}} \frac{\mu{\rm K}}{99.27~{\rm Jy}~{\rm sr}^{-1}}. 
\end{eqnarray}
The physical constants here are the Planck constant $h$, speed-of-light $c$, Boltzmann constant $k_{\rm B}$, and CMB mean temperature $T$.

We use two models for $dN/dS$, to be sure that our conclusions about the contribution of shot noise to the trispectrum are robust. The first (``model 1" hereafter)  is given in Ref. \cite{cmbpol_lensing_sources}, and is parameterized by $dN/dS=N_{0}/S^{\beta}$, where $\beta=2.15$, and $N_{0}=12~{\rm Jy}^{1.15}~{\rm sr}^{-1}$. The second (``model 2" hereafter), is obtained from tables of source counts in Ref. \cite{dezotti}. We fit a power law of the form used above to these tables, and obtain $\beta=2.74$ and $N_{0}=44.2$ for WMAP Q-band source counts. For the V-band source counts, we obtain $\beta=2.59$ and $N_{0}=21.8$. W-band source counts are not given in these tables, but we use the Q and V-band fits (cases W$\alpha$ and W$\beta$) as a model for W-band source counts. Flux cuts and conversion functions $g(x)$ appropriate for the W band, however, are used for the W-band trispectrum estimates.

We now use Eq.~(\ref{trispec_ps}) to obtain estimates for the unresolved point-source trispectra. We use a flux cut of $S_{c}\simeq 1.00~{\rm Jy}$, as it was found in Ref. \cite{wmap_1_ng} that this value reproduces observed WMAP point-source bispectra. 

For the Q-band, we obtain $\left \langle S^{4}\right \rangle =7.03\times 10^{-7}\mu{\rm K}^{4}$ and $\left \langle S^{4}\right \rangle =3.29\times 10^{-6}\mu{\rm K}^{4}$. For V-band data with model $1$, we obtain $\left \langle S^{4}\right \rangle =3.60\times 10^{-8}\mu{\rm K}^{4}$. For V-band data with model $2$, we obtain $\left \langle S^{4}\right \rangle =7.71\times 10^{-8}\mu{\rm K}^{4}$. For W-band data with model 1, we obtain $\left \langle S^{4}\right \rangle =1.89\times 10^{-9}\mu{\rm K}^{4}$. For model 2, case W$\alpha$, we obtain $\left \langle S^{4}\right \rangle =8.78\times 10^{-9}\mu{\rm K}^{4}$. For model 2, case W$\beta$, we obtain $\left \langle S^{4}\right \rangle =4.05\times 10^{-9}\mu{\rm K}^{4}$.

We plot a pessimistic estimate of the point source contamination in Fig.~\ref{fig:cldd_sources} for the WMAP V-band. As we know from the preceding discussion, the W-band contribution is negligible in comparison. The Q-band has the most worrisome amplitude of possible point source bias, however it is still more than an order of magnitude below the scale-invariant power spectrum shape at the low ($L<10$) multipoles and amplitudes ($\hat{A} < 5\times 10^{-3}$) which we probe most sensitively. We do not see evidence for any departure from zero in our model-independent $C_L^{\Delta \Delta}$ measurement, which would be characteristic of point source contamination.

\end{document}